\documentclass{article}[12pt]
\usepackage{amsmath,amssymb,amsthm,bm}
\usepackage{caption,graphicx}
\usepackage[all]{xy}

\usepackage{amssymb}
\usepackage{amsmath,amsthm}
\usepackage{latexsym}
\usepackage{amscd}
\usepackage{psfrag}
\usepackage{graphicx}
\usepackage[latin1]{inputenc}

\theoremstyle{plain}

\def\be{\begin{equation}}
\def\ee{\end{equation}}

\title{Theory of Morphogenesis}

\begin{document}

\maketitle

\begin{abstract} {\noindent A model of morphogenesis is proposed based upon seven explicit postulates.  The mathematical import and biological significance 
of the postulates are explored and discussed.\\\\
\centerline {Th\'eorie de la morphogen\`ese}\noindent{ R\'esum\'e  :}
Un mod\`ele de morphogen\`ese est propos\'e sur la base de sept postulats explicites. L'importance math\'ematique et la signification biologique
de ces  postulats sont explor\'ees et discut\'ees.}
\end{abstract}

\bigskip

\bigskip

\author{\noindent Andrey Minarsky,  National Research Academic University, Russian Academy of Science, Khlopina 8-3,194021, St.-Petersburg, Russia, tel. +7 812 232 0377 \\\texttt{minarsky@school.ioffe.ru}\\\\
\
Nadya Morozova$^*$, Institut des Hautes \'Etudes Scientifique, Le Bois-Marie, 35 route de Chartres, 91440, Bures-sur-Yvette, France, and Centre National de la Recherche Scientifique, tel.
+33 01 60 92 01 89\\
\texttt{morozova@ihes.fr}\\\\
\
Robert Penner$^*$,  Institut des Hautes \'Etudes Scientifique, Le Bois-Marie, 35 route de Chartres, 91440, Bures-sur-Yvette, France,  tel. +33 07 89 73 76 41\\ \texttt{rpenner@ihes.fr} \\\\
\
Christophe Soul\'e, Institut des Hautes \'Etudes Scientifique, Le Bois-Marie, 35 route de Chartres, 91440, Bures-sur-Yvette, France, and Centre National de la Recherche Scientifique,
tel. +33 01 60 92 66 21\\
\texttt{soule@ihes.fr}\\
}

\noindent \thanks{$^*$ Corresponding authors\\

\newpage

\section*{Introduction/Background}\label{sec:intro}

Morphogenesis is the evolution of shape of an organism together with the differentiation of its parts. 
The discovery of differential gene expression, that is, the spatio-temporal distribution of gene expression patterns during morphogenesis together with its key regulators, which are again given by gene expression, is one of the main recent achievements in developmental biology; see
Gilbert (2000) 
%\cite{G} 
and references therein. Nevertheless, differential gene expression cannot explain the development of the precise geometry of an organism and its parts; see
%\cite{L,MS}.
Levin (2012) and Morozova and Shubin (2012).

 The popular theory of morphogen gradients governing morphogenesis and accordingly differential gene expression,  though correct for some special cases, still leaves more 
 questions than answers, cf.\
 %\cite{W}. 
 Wolpert (2016). 
 For example, the mechanism of coordination of proper locations of specific morphogen production, the exact molecular pathways leading to morphogen gradient formation, the dependence of tissue formation and especially of  their geometrical shapes on exact gradients, along with many other key points, must still be elucidated in order to accept this theory as the basis for pattern formation rather than a part of molecular instruments implementing more general laws.  It is these more general laws which we shall postulate.

It is appealing to suggest the existence of a cell-surface molecular code which bears information about the geometrical pattern of an organism and thus coordinates the cascades of molecular events implementing pattern formation, e.g., differential gene expression, directed protein traffic, growth of microtubules and others.  Whatever the precise nature, this coding is epigenetic--literally, beyond genes--since diverse cell lineages with their diverse cell fates and morphogenetic evolutions nevertheless share the common genome of the organism itself.  Of course, the cell surface of an ovule is inherited as well as its DNA content.

This cell-surface location of code affords the possibility of involvement in signal transduction pathways. For example, received 
extra-cellular signals go to the nucleus or Golgi apparatus and influence the expression of specific sets of genes or the flow of protein traffic.  
Furthermore, this cell-surface location evidently mediates direct cell-to-cell interaction.  A set of experimental data
moreover confirms the significance of cell-surface information for pattern formation; see
%\cite{MS}
Morozova and Shubin (2012) and the references therein.

Though the concrete signal transduction pathways connecting the morphogenetic coding information and expression of given sets of genes are not yet elucidated, we can suggest a set of postulates and possible approaches for discovering the correspondence between this code and its realization in the given geometry of an organism in space-time.
This paper is a sequel to Morozova and Shubin (2012) and Morozova and Penner (2015)
%\cite{MP,MS} 
with the main new innovations being the inclusion of cell-to-cell communication, the emphasis on the role of cell potency, and the process of cell de-differentiation.

Our goal is to formalize the mechanisms and details of morphogenesis in order to uncover its underlying general laws,
two significant manifestations being embryonic development and physiological response to various crises such as
amputation, transplantation or biochemical intervention.

Let us finally note that our basic model of interacting cells with cell-surface coding quite naturally applies
{\sl mutatis mutandi} to other problems in theoretical biology.  For example in modeling the behavior of ant colonies, where there
is the paradigm that the colony is itself like an individual organism, it is well-known 
%\cite{b1,b2} 
(see  Cuvier-Hot et al. (2001) and (2005)) that individual ants carry
a carapace-surface code of esters by which they identify one another and propagate ant-to-ant signaling, thus capturing
at least this limited aspect of our detailed model of morphogenesis.

It is a pleasure to thank our colleague Minus van Baalen for excellent input.

\section{Overview/Fundamental Hypotheses}\label{sec:fundamental}

Our proposed theory of morphogenesis is based upon several fundamental hypotheses as follows:

\begin{enumerate}

\item[--]  For each cell in an organism there is a cell-surface distribution of chemical substances called its {\it (epigenetic) spectrum} governing morphogenesis.

\item[--] There is transmission to a certain collection of neighbors from each cell of its own epigenetic spectrum called {\it cellular signaling}.

\item[--] Each cell comprising an organism performs one of several possible {\it cell events} at various times, namely,
change of spectrum, change of position, change of shape including growth, mitotic division, and apoptosis (that is, programmed cell death).

%\item[--] Each cell in an organism furthermore carries two natural integer temporal descriptors: the number of cell divisions, called the {\it timer} $t$ from the zygote (that is, the fertilized egg), and the number of cell events, called the {\it stopwatch} $s$, from the most recent cell division leading to the cell;

%\item[--]called the {\it tree code} $\tau$; the tuple $C(c)=(A,t,s,\tau)$ is called the {\it state} of the cell $c$;

\item[--] There is a collection of {\it universal rules} obeyed throughout Nature for a specific cell event 
for each cell at each instant depending upon its own epigenetic spectrum and the cellular signals it receives.

\item[--] For each zygote for each organism, there are {\it optimal sequences} of cell events following
the universal rules which describe the normal evolution of the embryo.

\item[--] If an optimal cell event is impossible due for instance to crisis or malfunction, then the cell response is to 
{\it de-differentiate} and return its spectrum to that of its ancestor cell.

\item[--] The strength of the signal transmitted by a cell is inversely propotional to its {\it potency}.

\end{enumerate}

\medskip

More explicitly and to fix ideas, we assume that the spectrum is comprised of a collection of oligosacharide residues of glycoconjugates
lying in the lipid bilayer cell-surface membrane of each cell.  The concentration of these residues in different sectors of the cell can be described by a 
matrix with integer entries. 
Note that the spectrum of each cell could as well consist of other cell-surface molecules, in which case our general framework still applies.
Even short words (oligosaccharides) in the 6-12 letter alphabet of monosaccharides already provide ample combinatorial complexity.

\smallskip

The precise nature of signaling between cells likewise can remain unspecified.  The detailed signal could be a direct mechanical interaction of
cell-surface compounds or structures, or it could be molecular, such as ion exchange, ligand-receptor interactions or others
%this
including even potentially long-range chemical intercommunication.
We assume that
whatever its nature, the sent signal is itself determined by the spectrum of the sender cell which we may therefore take to be the signal itself.
It is in the {\sl interpretation} of that sent signal by the receiver cell that distinctions are made depending on precise details.  

\smallskip

Each cell in fact receives a set of signals from a collection of its neighbors, though again these ``neighbors'' may not be spatially proximate, and from these various signals determines an appropriate new {\it target spectrum}.  Under normal development after the cell event, the new spectrum agrees with the target spectrum, thus explicating a basic universal rule of morphogenesis.  Other external attributes of the cell, such as its position within the embryo or its shape may also alter as the result of a cell event.
If the target spectrum is unachievable by the cell in its current state, then the spectrum of the cell reverts to that of the previous cell event, a kind of backtracking 
which assumes a certain level of redundancy in the epigenetic spectrum.
In fact, much of the cell-surface molecular distribution of the two daughters of a divided cell
are simply inherited directly from the cell surface of the mother, while the region adjacent
to the division plane of each daughter must be filled in by certain universal rules according to the fourth postulate, so the code is indeed highly redundant across generations.

\smallskip

Notice that the plural is used in postulating optimal sequence{\bf s} of cell events.  This reflects actual data in
%\cite{us}
Bessonov et al.\ (2017) 
comparing different embryos of common species.  
It is not clear whether these competing
optimal responses should be regarded as a few discrete possible outcomes or perhaps in a more distributional sense.  At the same time and in the same way, the target spectrum must be taken in a probabilistic or distributional sense evidently mediated by code redundancy in any case.

\smallskip

Also notice the following immediate consequence of the postulates.  For a fully differentiated adult cell, there must be what we shall call a {\it harmony} between spectrum and received signal in order to determine equilibrium.  It is clear how this kind of harmony can determine shape: imagine excising
cells so as to remove their signals and drive the remaining cells to achieve the original harmonious shape.
We shall comment further on this harmony as a driving force for
morphogenesis.

\smallskip

The last postulate that signal strength varies inversely to the potency is the most difficult to explain here precisely because the concept of ``potency'' requires a number of further 
considerations.  This final postulate is explained further in $\S$\ref{sec:signal}. Roughly, the potency of a cell is its ability to produce a diversity of different cell states during an optimal sequence
of cell events discussed in $\S$\ref{sec:state}.  For instance, a zygote has maximal potency (totipotency), and a fully differentiated cell that admits no further mitotic divisions, such as a mammalian brain cell or eye lens cell, has minimal or no potency.  All cells at early stages of embryonic development, which are called {embryonic stem cells}, enjoy so-called pluripotency, or in some cases even the totipotency of the zygote, which allows them to alter their cell fates, while stem cells existing in tissues of adult organisms, or adult stem cells, enjoy  bipotency, meaning that they can produce only two types of cells--themselves and the differentiated cells of the corresponding tissue.

\smallskip

We hope and expect that future laboratory experimental work will confirm or refine aspects of the theory presented here.   Furthermore, a model based upon explicit incarnation of our postulates is 
currently being probed via computer implementation and experimentation 
%\cite{us}
in Bessonov et al.\ (2017).

%%%%
\section{Shapes}\label{sec:shape}

It is problematic to rigorously define the notion of {\it shape} or {\it form} in biology, 
%cf.\ \cite[9.1.1]{T}. 
cf.\ Section 9.1.1 of Thom (1983).
We shall do so at two scales: the microscopic shape of a cell and the macroscopic shape of an organism.

\smallskip

To define the {\it shape of a cell}\/~we proceed following
%cf.\ \cite{MP} 
Morozova and Penner (2015).   We assume that the cell $c$ contains
a distinguished point $O=O_c$ with respect to which it is star-convex, that is, the line segment $\overline{OP}$
lies in $c$ for any other point $P$ in $c$.
Though the example of a neuron cell shows this is not strictly true for all cells, we can accept that the few such counter-examples are not especially critical in determining shapes of organisms. Specifically, we take $O$ to be the so-called microtubule organizing center or centrosome.  Insofar as the tubules that emanate from the centrosome keep in place the cell surface lipid bilayer, this further mathematical assumption that cells are star-convex with respect to the centrosome is biologically sound.

\smallskip

It follows that the shape in space of the cell membrane $\mu(c)$ of $c$ can be described by a positive real function $\sigma=\sigma_c : S^2 \to {\mathbb R}_{ \, > \, 0}$ 
on the unit-radius two-dimensional sphere $S^2$ centered at $O$, namely, if $P_0 \in S^2$, then the point $P \in \mu (c)$ in the direction of $P_0$ 
from $O$ is uniquely defined by the equality 
$
\overset{\longrightarrow}{OP} = \sigma \, (P_0) \ \overset{\longrightarrow}{OP_0}$ of vectors.
Thus as a subset of Euclidean space ${\mathbb R}^3$, the cell $c$ in space is given by the convex set
$B_{\sigma}(O)=\{ P\in{\mathbb R}^3:||\overset{\longrightarrow}{OP}||\leq \sigma(P_0)\}\subset{\mathbb R}^3$ containing $O$.
To fix ideas, let us assume that $\sigma_c \in L^2(S^2)$, i.e., $\sigma_c$ is square integrable for each cell $c$.

\smallskip

Now turning to the {\it shape of an organism} $\Omega$ regarded as the union of its constituent cells, one encounters the following difficulty emphasized by Ren\'e Thom {\sl loc.\ cit.}:
At an instant in time the organism $\Omega$ is embedded as a closed subset in ${\mathbb R}^3$, and the coordinate axes can be chosen to coincide with the three embryonic axes (anterior-posterior, dorsal-ventral, left-right) of the organism determined already in the zygote. Insofar as the organism $\Omega$ can move in space, it admits multiple manifestations as subsets, and it is not clear how to specify precisely when two such explicit manifestations of $\Omega$ are nearby.

\smallskip

We propose to proceed as follows. The shape of $\Omega$ is determined by a finite and connected graph $I(\Omega)$, called the {\it graph of adjacency} whose vertices are given by the cells of $\Omega$ with an edge between vertices $c_1$ and $c_2$ when the cells $c_1$ and $c_2$ touch one another; in fact,
$I(\Omega)$ is equipped with a natural metric assigning to the edge between $c_1$ and $c_2$ the distance between $O_{c_1}$ and $O_{c_2}$ in ${\mathbb R}^3$, in contrast to the simpler combinatorial length determined by the number of edges traversed.  This metrized graph $I(\Omega)$ can be isometrically embedded in ${\mathbb R}^3$ in such a way that the vertex $c$ is mapped to the distinguished point of the cell $c$, and  $\sigma_c$ determines the extent of the cell in space. Notice that the collection of functions $\sigma_c$ are not arbitrary, e.g., because two cells cannot overlap. A crucial point is that we {\sl do not fix the embedding} of $I(\Omega)$ into ${\mathbb R}^3$. Two different closed subsets of ${\mathbb R}^3$ have the same shape when they share the same data $(I(\Omega) , \sigma_c)$.

\smallskip

We can now go further and give a notion of distance between two organisms $\Omega_1$ and $\Omega_2$. Namely we may take the {\it Gromov-Hausdorff (GH) distance} 
%(cf.\ \cite{GH1})  
(see\ Gromov (2000)) between the metrized graphs $I(\Omega_1)$ and $I(\Omega_2)$.
There is also the related notion considered in 
%\cite{MP} 
Morozova and Penner (2015) where one regards the organism $\Omega=\cup_{c\in\Omega} B_{\sigma_c}(O_c)\subset{\mathbb R}^3$ as the union of its cells in space as a metric subspace of ${\mathbb R}^3$ and again measures distances between organisms using GH distance.  

It is well-accepted,
%\cite{GH2,GH3} 
cf.\ Bronstein et al.\ (2006) and M\`emoli and Sapiro (2005),
that GH distance is of utility in comparing objects moving in space-time,
so this suitably solves Thom's problem of moving organisms in any case.  However, the computational
complexity of calculating GH distances is likewise well-known; see
%\cite{GH}. 
M\`emoli (2007). Let us also here distinguish G from H: The H distance
(GH distance, respectively) is defined on pairs of metric spaces (pairs of measured metric spaces, respectively).  Internal cell
contents could in prinicple be added to our basic model via suitable measures on adjacency graphs.

%%%
\section{Cell State and Cell Event}\label{sec:state}

As discussed in $\S$\ref{sec:fundamental} we postulate that the development of an organism is driven by {cell-surface molecular codes} called (epigenetic)
spectra of its constituent cells.  As a discrete approximation of this code, we consider the set $Mat$ of $N$-by-8 matrices $A_c=(a_{ij})$ with natural integer entries $a_{ij}$, where $N$ is the number of species of glycoresidues we shall record for each cell, and the three coordinate planes decompose each cell surface into eight orthants within which we record the number $a_{ij}\geq 0$ of each of the $N$ species, for $i=1,\ldots ,N$ and $j=1,\ldots,8$.  This is just a crude simplification.  As in 
%\cite{MP} 
Morozova and Penner (2015), a more sophisticated approach would be to record the actual densities with further real-valued functions defined on the sphere $S^2$, one such function for each species for each cell, rather than the discrete model with integral matrices considered here.

\smallskip

There are several data intrinsically associated with each cell $c$, namely,
\begin{enumerate}
\item[--] the epigenetic spectrum $A_c\in Mat$,
\item[--] the shape function $\sigma_c\in L^2(S^2)$,
\item[--] the coordinates of the distinguished point $O_c\in {\mathbb R}^3$,
\item[--] the number $t_c$ of cell divisions directly leading to $c$ from the zygote called the {\it cell timer},
\item[--] the number $s_c$ of cell events occurring since the most recent cell division called the {\it cell stopwatch},
\item[--] the relative age of the most recently inherited centrosome, $\alpha_c=m$ for the older (mother) and $\alpha_c=d$ for the younger (daughter) centrosome, called the {\it $m/d$ invariant}.
\end{enumerate}

These data are intrinsic in the sense that the cell might be removed from its organism yet preserving each of these
attributes which could then be measured.  
Together these data comprise the {\it cell state} $S_c=(A_c,\sigma_c,O_c,t_c,s_c,\alpha_c)$, and we shall regard $I_c=(A_c,t_c,s_c,\alpha_c)$ as the {\it internal state}
and $E_c=(\sigma_c,O_c)$ as the {\it external state} of the cell $c$.
We should note that the first three pieces of data in $S_c$ can be organized into a  {bundle over the configuration space} of distinguished points in space with fiber given by shapes and spectra
%, cf.\ \cite{MP}
as in Morozova and Penner (2015).

The biological and mathematical significance of $A_c$, $\sigma_c$ and $O_c$ have already been discussed.  In order to elucidate the two
timers, let us first construct the {\it tree $T=T_\Omega$ of cell events} whose vertices are in correspondence with the cell states $S_c$ with an edge connecting
vertices when they are related by a cell event.   The zygote in its initial state forms the root of the tree $T$ and has valence 2 corresponding to the
fact that it divides from its current state at the outset of the construction; other 2-valent vertices arise from change of spectrum, position or shape, while 3-valent vertices correspond to division and 1-valent vertices to apoptosis.  The tree $T$ is metrized where the length of an edge is given by the temporal duration of the corresponding cell event.    

The path in $T$ from the zygote to the vertex of $T$ labeled by cell state $S_c$ passes through a certain number of 3-valent vertices, and this number is the value of the timer $t_c$.  The biological determination of $t_c$ can be approximated in terms of the length deficit of the so-called telomeric tail
%this
of the DNA contained in the cell $c$, which loses one telomere for each division, 
%cf.\ \cite{A,G}; 
cf.\ Alberts et al. (2002) and Gilbert (2000);
strictly speaking, a single cell division might remove several telomeres from the tail due to oxidative effects, and indeed there are proteins called TERTs which serve to lengthen the telomeric tail, 
%cf.\ \cite{S};
cf.\ Shampay and Blackburn (1988); let us nevertheless regard $t_c$ as an intrinsic datum roughly determined by the telomeric tail length and given precisely by this and some other intrinsic cell data which can remain unspecified for now. 

Analogously, the path in $T$ from zygote to the vertex $v$ has a last passage through a 3-valent vertex before arrival at $v$, and the 
number of 2-valent vertices it meets after visiting this 3-valent vertex, or in other words the number of changes of spectrum, shape or position that occur 
from the most recent division, gives the cell stopwatch $s_c$ in terms of $T$.  The biological determination of the stopwatch requires a short digression as follows.
All cells contain microtubules in particular supporting the cell surface, as we have mentioned, and have a specific microtubule organizing center which gives 
a distinguished point within each cell.  In fact, microtubules are not static and cycle through a process of adding to the base (proximal) and removing from the tip 
(distal), and this cycle time in fact correlates with the cell cycle controlling mitosis. Thus, a notch on the microtubule moves up and away from the base towards the tip, and the distance of this notch from the base again gives an approximate biological interpretation to the intrinsic stopwatch $s_c$.  

To explain the {\it m/d} invariant $\alpha=\alpha_c$ let us note that the centrosome is duplicated during mitotic as well as meiotic cell division; see
%cf.\ \cite{A,G}.  
Alberts et al. (2002) and Gilbert (2000).
The daughter cell inheriting the older centrosome has $\alpha=m$, and the other daughter cell has its $\alpha=d$.
For diplosomes whose centrosome is comprised of two centrioles and which includes all animal cells, one centriole is older than the other and each duplicates to produce another pair of complete centrosomes each comprised of two centrioles; the daughter cell inheriting the oldest of the four constituent centrioles has $\alpha=m$ the other having $\alpha=d$.  The $m/d$ invariant is indeed intrinsic in particular for diplosomes insofar as asymmetries between $m$ and $d$ centrioles go beyond simply age presenting notable differences in molecular composition, function and ultrastructure.  The ovum and sperm in the diplosomic case each contain just one centriole, and the former is $m$ in the zygote.  Numerous experiments for diplosomes have shown that cell fate is tied
to the $m/d$ invariant, 
%cf.\ \cite{RG} 
cf.\ Reina and Gonzalez (2014) and the references therein.  We again assume this $m/d$ invariant is likewise intrinsic in general by these or other unspecified attributes.

Keeping track of the $m/d$ invariant starting from the zygote, each cell $c$ in an organism has a well-defined word of length $t_c$ in the letters $\{ m,d\}$ called 
the {\it $m/d$ code} which uniquely determines the phylogeny of its centrosome starting from the zygote.  It is an interesting question whether the full $m/d$ 
code is intrinsic or perhaps just a terminal segment of it of some fixed length definitely greater than or equal to one.  It seems unlikely that cell events could depend upon more that the last few letters since otherwise presumably inevitable errors in $m/d$ code would be catastrophic for embryogenesis.

Fix some organism $\Omega$ with zygote $z$ and consider a cell state $S_c$ labelling a vertex on the tree $T$ its optimal cell events with its well-defined subtree
$T(S_c)\subseteq T=T(S_z)$ with this vertex as its root.  Define the collection $X(S_c)$ of all pairs $(A_d,\alpha_d)$ occurring as data among vertices of 
$T(S_c)$.  The ratio of the measure of $X(S_c)$ to that of $X(S_z)$ for some appropriate measure of the set of all pairs comprised of spectrum and $m/d$ invariant is the {\it (normalized) potency} of the cell state $S_c$. 

Potency is {\sl not} an intrinsic attribute of the cell state in the sense discussed previously, and its definition requires {\sl a priori} choosing one particular
tree of optimal cell events.  Despite
much attention, we do not know a reasonable definition of intrinsic potency since one must specify under exactly which conditions
a cell in its state is allowed to evolve: under {\sl all possible conditions} being too broad and unmeasurable and under {\sl specific laboratory conditions} 
being too specialized.  Notice however that if the full $m/d$ code were intrinsic, then potency for counting measure on $Mat\times\{m,d\}$ could actually 
be determined in laboratory experiment without killing the organism: sample each type of internal cell state $(A,\alpha)$ in the complete mature organism and 
compare with the histogram of sampled
initial segments of $m/d$ code.

\section{Signaling and Cell Response}\label{sec:signal}
 
\noindent We have already in $\S$\ref{sec:fundamental} explained that each cell $c$ of an organism $\Omega$ provides its signal to a collection of its neighbors, which may be spatially non-proximate, and the signal is given by its own spectrum.  The set of cells of $\Omega$ that receives this signal can be defined in various ways.  For example, the signal might propagate uniformly in all directions or may have vectorial characteristic, it might decay with spatial distance from $c$ or with combinatorial distance in the graph of adjacency from the vertex corresponding to $c$, the simplest possibility, or it might depend on the subsets of epigenetic spectra through which it is transmitted.  

Each cell $c$ of $\Omega$ thus also receives a certain collection of signals from its neighbors, and these must be combined in some manner to
produce the target spectrum also discussed in $\S$\ref{sec:fundamental}.  There are again various possibilities ranging from a simple average over signals received possibly weighted by distance or other attributes again including perhaps the spectra through which it is transmitted, also possibly allowing for stochastic effects and depending inversely upon the potency of the sender in any case according to our final postulate.

In the optimal situation, the spectrum of the cell after the cell event coincides with the target spectrum.  In particular if the cell $c$ in its state with spectrum $A$ is provided with a target spectrum that agrees with $A$, this means that the optimal (coded) cell event for the cell $c$ is confirmed by the signal. It is this ``harmony'' between cell current spectrum and the target spectrum determined by the received signal that communicates to the cell that it should ``move'' along the optimal tree of cell events. 
In this way, the shape of the organism through signaling can communicate equilibrium to its constituent cells at the conclusion of morphogenetic processes as discussed in $\S$\ref{sec:fundamental}.

More generally though, the target spectrum differs from the current spectrum and the cell state evolves.  This evolution normally follows a
pattern of differentiation, by which we mean that the cell states becomes more and more specialized, less capable of diverse evolution, thus with
diminished potency.

Note that cell events depend on parameters. For instance the division of  a cell  requires the specification of a plane of division.   Another aspect of cell events requires determining the rules for the resulting distribution of coding species (epigenetic spectrum) on a cell surface after a cell event.  For example for the cell event of division, it is quite reasonable to postulate that half of the daughter cell-surface spectra are directly inherited unchanged from the mother spectrum,
one aspect of redundancy mentioned before, while the remaining daughter spectra adjacent to the division plane are filled in by certain rules as yet to be determined.

We have thus far concentrated primarily on normal evolution of an embryo and finally briefly consider cell response under unusual circumstances.  For example an amputated limb in the frog species {\it Xenopus} is capable of regeneration, and even small body fragments of the {\it Planaria} worm can generate an entire and complete organism, 
%cf.\ \cite{L}
cf.\ Levin (2012). Even human babies are capable of regenerating amputated fingertips during the first months following birth it turns out. The removal
of the cell membrane from a plant cell produces a so-called callus of many undifferentiated cells capable of generating an entire and complete plant organism. 
Transplantation of  limb fragments in non-native orientations in {\it Drosophilia}, {\it Axolotl}  and other species
can result in supernumerary limb regeneration as well as other bizarre outcomes. The literature abounds with experiments illustrating these remarkable phenomena, 
%cf.\ \cite{G, L} 
cf.\ Gilbert (2000) and Levin(2012)
and references therein.

In our model, the cell in its cell state is provided by signaling with a target spectrum and then responds with its optimal cell event under normal
conditions, but if the conditions are not normal so the target spectrum is for some reason unachievable, then we posit that the cell has the only possible responses of stagnation (that is, no cell event), cell death (a form of apoptosis under these unusual conditions), or a de-differentiation (that is, the return to its
previous epigenetic spectrum).  

In particular, the cell epigenetic spectrum can return to that of its mother in the case of cell division and may then de-differentiate further perhaps to its grandmother and beyond, or it may perhaps again divide.  This cycle of devolution to ancestor and division accurately reflects
the kind of de-differentiation and cell proliferation in a blastema that typically precedes regeneration in experiments.  

Consider the case of amputation with the limb stub first of all covered by this high potency 
de-differentated region $R$.  Now the process of regeneration can proceed with the 
differentiated tissue in the stub, of low potency and hence strong sent signal, adjacent to the
high-potency hence weak-signal cells in $R$.  Meanwhile, the cells in the stub were in harmony with the nearby cells in the organism before amputation, so the determination not only of future shape of $R$ but also future differentiation within $R$ is plausibly driven by signaling inversely proportional to potency. 

\medskip

We have articulated and discussed in this short paper
seven explicit laws of morphogenesis which when taken
together explain myriad phenomena among
experiments in the literature.


\begin{thebibliography}{999}
\makeatletter
\renewcommand\@biblabel[1]{}
\makeatother

\bibitem {A} Alberts, B.,  Johnson, A.,  Lewis, J., Raff, M., Roberts, K., and Walter, P., 2002.
{\it Molecular Biology of the Cell}, 4th Edition, Garland Publishing, New York, New York.

\bibitem{us} Bessonov, N., Butuzova,O., Minarsky,  A., Morozova, N., Penner, R., Soul\'e, C., and Tosenberger, A., 2017.
Computational tool for molecular code in morphogenesis.
{\it In preparation}.

\bibitem{GH2} Bronstein, A.M., Bronstein, M.M., and Kimmel, R., 2006. 
Efficient computation of isometry-invariant distances
between surfaces.
{\it SIAM J.\ Sci.\ Comp.}\ {28}, 1812-1836.

\bibitem{b1} Cuvillier-Hot, V., Cobb, M., Malosse, C., and Peeters, C., 2001.
Sex, age and ovarian activity affect cuticular hydrocarbons in Diacamma ceylonense, a queenless ant.
{\it J.\ Insect Physiol.} {47}, 485-493.

\bibitem{b2}
Cuvillier-Hot, V., Renault, and V., Peeters, C. 2005.
 Rapid modification in the olfactory signal of ants following a change in reproductive status.
 {\it Naturwiss.}\ {9}, 73-77.

\bibitem{G} S.\ Gilbert, S., 2000.
{\it Developmental biology}.
Sinauer Associates Inc., Sunderland, MA.

\bibitem{GH1} Gromov, M., 2007.
{\it Metric Structures for Riemannian and Non-Riemannian Spaces}.
Birkh\"auser, Basel, CH.

\bibitem{GH} M\'emoli, F., 2007.
On the use of Gromov-Hausdorff distances for shape comparison. {\it In}
Botsch, M. and Pajola, R., eds., {\it Eurographics Symposium on Point-Based Graphics}, 
AK Peters, Natick, MA.

\bibitem{GH3} M\'emoli, F., and Sapiro, G., 2005.
 A theoretical and computational framework for isometry invariant recognition
of point cloud data.
{\it Found.\ Comput.\ Math.}\ {5}, 313-347.

\bibitem{L} Levin, M., 2012.
Morphogenetic fields in embryogenesis, regeneration, and cancer: Non-local
control of complex patterning.
{\it BioSystems} {109}, 243-261. 
                 
\bibitem{MP} Morozova, N., and Penner, R., 2015.
Geometry of morphogenesis.
{\it In} Mondaini, R., ed., {\it BIOMAT 2014, Proceedings of the International Symposium on Mathematical and Computational Biology}, World Scientific, Singapore, 331-350.

\bibitem{MS} N.\ Morozova, N., and Shubin, M., 2012.
The geometry of morphogenesis and the morphogenetic field concept.
{\it In} 
Capasso, V.,  Gromov, M., Harel-Bellan, A., Morozova, N., and Pritchard, L.L., eds., {\it Pattern Formation in Morphogenesis-Problems and
Mathematical Issues}
{15}, Springer Verlag, Boston, 255-282.

\bibitem{RG} Reina, J., and Gonzalez, C. 2014.
When fate follows age: unequal centrosomes in asymmetric cell division.
{\it Philos.\ Trans.\ R.\ Soc.\ Lond. B Biol.\ Sci.}~369.

\bibitem{S} Shampay J., and Blackburn, E.H., 1988.
Generation of telomere-length heterogeneity in Saccharomyces cerevisiae.
{\it Proc.\ Natl.\ Acad.\ Sci.\ U.S.A.} {85}, 534-538.

\bibitem{T} Thom, R., 1983.
{\it Mathematical Models of Morphogenesis}.
Ellis Horwood, Hemstead, UK.

\bibitem{W} Wolpert, L., 2016.
Positional information and pattern formation.
{\it Curr.\ Top.\ Dev.\ Biol.}\ {117}, 597-608.

\end{thebibliography}
\end{document}